\documentclass[%
 aip,
 amsmath,amssymb,
 reprint,%
]{revtex4-1}
\usepackage{graphicx}
\usepackage{dcolumn}
\usepackage{bm}

\usepackage[utf8]{inputenc}
\usepackage[T1]{fontenc}
\usepackage{siunitx}
\usepackage{mathptmx}
\usepackage{etoolbox}

\makeatletter
\def\@email#1#2{%
 \endgroup
 \patchcmd{\titleblock@produce}
  {\frontmatter@RRAPformat}
  {\frontmatter@RRAPformat{\produce@RRAP{*#1\href{mailto:#2}{#2}}}\frontmatter@RRAPformat}
  {}{}
}%
\makeatother
\begin{document}

\preprint{AIP/123-QED}

\title[Fast Recovery Dynamics of GaSbBi-based SESAMs for high-fluence operation]{Fast Recovery Dynamics of GaSbBi-based SESAMs for high-fluence operation}

\author{Maximilian C. Schuchter}
\email{maximilian.schuchter@tuni.fi}
\affiliation{ 
Tampere University, Optoelectronics Research Centre (ORC), Physics Unit, FI-33720 Tampere, Finland}
\affiliation{ 
ETH Zurich, Department of Physics, Institute for Quantum Electronics, 8093 Zurich, Switzerland}

\author{Joonas Hilska}%
\affiliation{ 
Tampere University, Optoelectronics Research Centre (ORC), Physics Unit, FI-33720 Tampere, Finland}

\author{Markus Peil}
\affiliation{ 
Tampere University, Optoelectronics Research Centre (ORC), Physics Unit, FI-33720 Tampere, Finland}

\author{Eero Koivusalo}
\affiliation{ 
Tampere University, Optoelectronics Research Centre (ORC), Physics Unit, FI-33720 Tampere, Finland}
\affiliation{ 
Reflekron Ltd, Kauhakorvenkatu 53 B, Tampere 33710, Finland}

\author{Marco Gaulke}
\affiliation{ 
ETH Zurich, Department of Physics, Institute for Quantum Electronics, 8093 Zurich, Switzerland}

\author{Ursula Keller}
\affiliation{ 
ETH Zurich, Department of Physics, Institute for Quantum Electronics, 8093 Zurich, Switzerland}
\affiliation{ 
ETH Zurich, Department of Electrical Engineering and Information Technology, 8093 Zurich, Switzerland}

\author{Mircea Guina}
\affiliation{ 
Tampere University, Optoelectronics Research Centre (ORC), Physics Unit, FI-33720 Tampere, Finland}


\date{\today}

\begin{abstract}
Modelocked lasers operating at \si{2} – \SI{3}{\micro\meter} wavelength region are interesting for various spectroscopic applications. To this end, GaSb-based semiconductor saturable absorber mirrors (SESAMs) are developing fast as a practical technology for passive modelocking. Yet, such SESAMs suffer from either too high two-photon absorption or slow absorption recovery dynamics. This study introduces GaSbBi quantum wells (QWs) as a novel platform to ensure a larger material selection for engineering GaSb-based SESAMs with decreased two-photon absorption and ultrafast absorption recovery time. Three GaSbBi QW SESAM designs were fabricated to compare their performance against conventional GaInSb QW SESAMs. The first structure makes use of typical GaSb barriers and exhibits comparable characteristics to the conventional design, including a saturation fluence of \SI{1.09}{\micro\joule\per\square\centi\meter}, modulation depth of \SI{1.41}{\percent}, and a fast interband recovery time of \SI{6.03}{\pico\second}. The second design incorporated AlAs\textsubscript{0.08}Sb\textsubscript{0.92} barriers, achieving a reduced two-photon absorption, though at the cost of higher non-saturable losses due to unintended Bi droplet formation during growth of the AlAs\textsubscript{0.08}Sb\textsubscript{0.92}/GaSbBi QW heterostructure. Importantly, it maintained a fast interband recovery time (\SI{30}{\pico\second}), overcoming the slow recovery dynamics exhibited by standard GaInSb QW SESAMs with AlAs\textsubscript{0.08}Sb\textsubscript{0.92} barriers. The third design explored GaSbBi QWs with higher Bi content targeted for longer wavelength operation at \SI{2.3}{\micro\meter}, which exhibited fast recovery times and good nonlinear reflectivity characteristics. However, the higher Bi content resulted in elevated non-saturable losses.These results highlight the potential of GaSbBi QWs for SWIR SESAMs, opening the path for further epitaxial optimization to enhance their performance.
\end{abstract}

\maketitle

%

\phantomsection
\vspace{0.7cm}

The SWIR spectral region (1.4 – \SI{3}{\micro\meter}) has become increasingly important in multiple research areas in photonics, biomedicine, and spectroscopy \cite{zhang_advances_2024,seddon_mid-infrared_2013, isensee_biomedical_2018}. Many of these research areas rely on ultrafast laser technologies, for which Semiconductor Saturable Absorber Mirrors (SESAMs) have become an established technology for passive modelocking. In the SWIR regime, the most developed SESAMs are based on quantum well (QW) heterostructures grown on GaSb substrate. The first GaInSb/GaSb QW SESAMs at \SI{2}{\micro\meter} exhibited an unusual ultrafast recovery time for as-grown structures\cite{paajaste_absorption_2014, paajaste_gasb-based_2012, heidrich_full_2021, alaydin_bandgap_2022}, which was later confirmed and fully characterized also for SESAMs operating at \SI{2.4}{\micro\meter} \cite{heidrich_full_2021, alaydin_bandgap_2022}. These structures have been successfully used to modelock a large variety of lasers, including VECSELs operating at \SI{2}{\micro\meter} \cite{harkonen_picosecond_2010, heidrich_324-fs_2022, gaulke_optically_2024}, as well as solid-state lasers at \SI{2.3}{\micro\meter}, such as in \cite{barh_watt-level_2021, tyazhev_upconversion-pumped_2024}.

While these conventional GaSb-based SESAMs exhibit adequate modulation depths, low nonsaturable losses, favorable recovery times, and supported successful modelocking, they suffer from a small rollover parameter (\(F_2\)), primarily attributed to two-photon absorption (TPA) effects, which limit pulse shortening in modelocked operation \cite{heidrich_full_2021}. Following on the development path for near-infrared SESAMs, GaInSb/AlAs\textsubscript{0.08}Sb\textsubscript{0.92} SESAMs exhibit higher \(F_2\) parameters due to reduced TPA in barriers with higher bandgap energy (\SI{1.639}{\electronvolt}). Materials with a bandgap exceeding \SI{1.2}{\electronvolt} have negligible TPA at the operation wavelength of \SI{2060}{\nano\meter} (i.e. \SI{0.6}{\electronvolt} photon energy). A recent study \cite{schuchter_composition-controlled_2025} proved this assumption, and investigations also revealed the influence of the strain of the absorbing QWs. While the AlAs\textsubscript{0.08}Sb\textsubscript{0.92} barriers effectively reduced TPA owing to their high bandgap energy, they induce notably slow recovery times (\textgreater \SI{500}{\pico\second}), rendering GaInSb/AlAs\textsubscript{0.08}Sb\textsubscript{0.92} SESAMs unsuitable for modelocking applications. Furthermore, at longer wavelengths beyond \SI{2.4}{\micro\meter}, the use of ternary GaInSb QW material becomes impractical owing to strain limitations. Extending the wavelength requires either increasing the In composition or using thicker QWs, which eventually causes lattice relaxation, possibly compromising the structural integrity of the SESAM \cite{alaydin_bandgap_2022}. When utilizing InGaAsSb QW SESAMs with less strain, the study \cite{schuchter_composition-controlled_2025} indicated that the problem of slow recovery times persisted.

A novel solution could be provided by using GaSbBi QWs as absorbers. The incorporation of bismuth (Bi) into GaSb results in substantial bandgap reduction (\SI{30}{\milli\electronvolt} – \SI{35}{\milli\electronvolt} per \%Bi \cite{kopaczek_temperature_2013}) and induces much less strain per bandgap shift than GaInSb (\SI{1160}{\milli\electronvolt} per 1\% strain \cite{kopaczek_low-_2014} vs. \SI{140}{\milli\electronvolt} per 1\% strain \cite{vurgaftman_band_2001}), making GaSbBi an attractive QW absorber alternative for SESAMs. Secondly, GaSbBi has a larger valence band offset and thus maintains the type-I band alignment with respect to GaSb \cite{yue_molecular_2018} at longer wavelengths. While these characteristics provide a pathway for optimizing SESAM performance at longer wavelengths, we will also show that they solve the issue of slow recovery times associated with using AlAs\textsubscript{0.08}Sb\textsubscript{0.92} barriers. 

Previous studies on GaSbBi alloys have focused on optimizing the molecular beam epitaxy (MBE) growth and investigating structural and photoluminescence properties \cite{delorme_molecular_2017, hilska_epitaxial_2019,yue_molecular_2018,rajpalke_high_2014,rajpalke_bi_2015,gladysiewicz_electronic_2016}, including the radiative lifetimes \cite{smolka_influence_2023,rogowicz_carrier_2022}. The necessary MBE growth conditions for high Bi incorporation (up to ~\SI{14}{\percent} Bi \cite{delorme_molecular_2017}) are generally understood \cite{hilska_epitaxial_2019}; however, despite their promising characteristics for SWIR optoelectronics, no SESAM or VECSEL devices utilizing GaSbBi alloys have been realized to date. This study reports on the fabrication of GaSbBi-based SESAMs and their characteristics at the \si{2} – \SI{2.3}{\micro\meter} SWIR wavelength range.

\phantomsection
\vspace{0.5cm}

\begin{figure}
\includegraphics{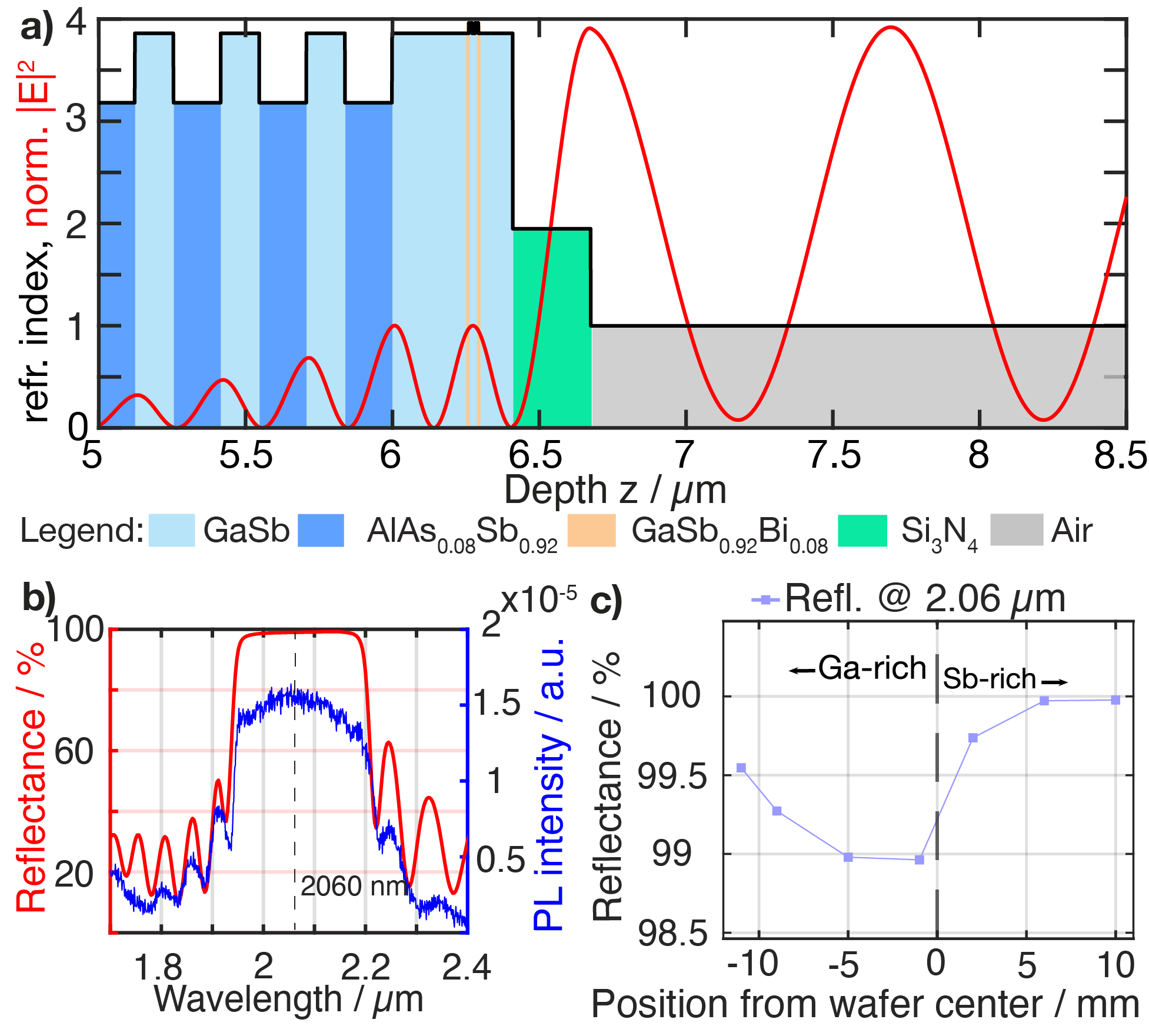}
\caption{\label{fig:Figure1} SESAM \#1 design: a) A partially shown, \num{21.5} pair DBR of \(\lambda/4\) pairs of GaSb (light blue) and AlAs\textsubscript{0.08}Sb\textsubscript{0.92} (dark blue), followed by two GaSb\textsubscript{0.92}Bi\textsubscript{0.08} QWs, shown in orange, placed in the antinode of the electric field intensity, shown in red. A \(\lambda/4\) Si\textsubscript{3}N\textsubscript{4} AR coating was PECVD deposited on top. The x-axis denotes the growth depth \(z\). b) Linear reflectance and room temperature photoluminescence. The measurements were performed at the wafer center with optimal Bi incorporation. c) The peak reflectance at the design wavelength of \SI{2060}{\nano\meter} measured over the wafer.
}
\end{figure}

The SESAMs were grown on 2" (100) n-GaSb substrates using molecular beam epitaxy (MBE) with conventional effusion cells for Ga, Al, and Bi, and valved cracker sources for As and Sb. The design of SESAM \#1 is shown in Figure \ref{fig:Figure1}a), wherein a distributed Bragg reflector (DBR) consisting of \num{21.5} pairs of quarter-wavelength low refractive index lattice-matched AlAs\textsubscript{0.08}Sb\textsubscript{0.92} and high refractive index GaSb layers was used. The DBR exhibits a reflectivity around \SI{99.9}{\percent} at the design wavelength of \SI{2060}{\nano\meter}. The DBR was followed by a GaSb layer incorporating two 10-\si{\nano\meter} thick GaSbBi QWs placed at the antinode of the standing wave electric field, as shown in Figure \ref{fig:Figure1}. The QWs contain approximately \SI{8}{\percent} Bi and were separated by a \SI{20}{\nano\meter} thick GaSb barrier. The GaSbBi QWs and the GaSb barrier were grown stationarily \cite{hilska_epitaxial_2019}, while the other layers were grown with normal substrate rotation. The stationary growth resulted in a well-controlled gradient in the flux distributions across the wafer, thus inducing a gradient in the Bi incorporation. This allowed access to a large range of compositions to understand the impact of Bi incorporation on the optical quality of the QWs over the wafer. 

For SESAM \#2, the structure was grown identically to SESAM \#1, however, the cavity barrier material was changed from GaSb to AlAs\textsubscript{0.08}Sb\textsubscript{0.92}. For SESAM \#3, the Bi content in the QWs was increased from approximately \SI{8}{\percent} Bi to \SI{13}{\percent} Bi to extend the operation wavelength to \SI{2.3}{\micro\meter}, while keeping the barrier material as GaSb. For SESAM \#3, the thicknesses of all layers were adjusted based on the design wavelength to ensure the QWs were still placed at the antinodes. 

The differences between the structural design of the three SESAMs are summarized in Table \ref{tab:table1}. In addition, the SESAMs were anti-reflection (AR) coated with a quarter-wavelength Si\textsubscript{3}N\textsubscript{4} layer deposited by plasma-enhanced chemical vapor deposition (PECVD) to enhance QW absorbance. 

All the as-grown SESAMs were characterized by measuring the reflectance and photoluminescence (PL). Later, the wafer material with the ideal Bi incorporation regime was chosen for nonlinear SESAM characterization. Figure \ref{fig:Figure1}b) shows the measured reflectance and room temperature photoluminescence (PL) spectrum of SESAM \#1 at the optimal Bi incorporation position near the center of the wafer. The reflectance is near \SI{100}{\percent}, illustrating the high reflectivity of the DBR together with low absorbance of the GaSbBi QWs. The PL spectrum is very broad, essentially covering the full stop band of the DBR. Similarly broad PL emission at room temperature has been observed for bulk GaSbBi epilayers \cite{delorme_molecular_2017}. 

Figure \ref{fig:Figure1}c) shows the measured reflectance at the SESAM-design wavelength of \SI{2060}{\nano\meter} measured over the wafer along the maximal Sb/Ga gradient \cite{hilska_epitaxial_2019}. At the negative coordinates of Figure \ref{fig:Figure1}c), the Sb/Ga ratio is below stoichiometric (Sb/Ga < 1). While we still see some absorbance (i.e., non-\SI{100}{\percent} reflectance), which is consistent with some Bi incorporation in the QWs \cite{hilska_epitaxial_2019}, it is not the ideal incorporation. In contrast, at the center of the wafer, corresponding to the stoichiometric Sb/Ga ratio (Sb/Ga \(\approx 1\)), the absorbance increases, forming the ideal Bi incorporation regime. For the positive coordinates, the Sb/Ga ratio becomes above stoichiometric (Sb/Ga > 1), leading to very little Bi incorporation in the QWs \cite{hilska_epitaxial_2019}, and thus no absorbance at the operation wavelength. Here, the reflectance essentially reaches \SI{100}{\percent}. 

Nonlinear reflectivity and nonlinear recovery dynamics were measured for selected SESAMs, i.e., components from wafer regions with good material quality. For this, we used the dedicated setups described in detail in \cite{heidrich_full_2021}. 

For fitting the temporal response of the SESAM, only the Shockley-Read-Hall recombination with the conventional bi-exponential decay function \cite{maas_growth_2008} is considered. Therefore, two time constants, \(\tau_1\) and \(\tau_2\), are obtained, with a weighting factor \(A\) between them. The first time constant, \(\tau_1\), describes the fast decay based on intraband thermalization, usually on the order of hundreds of femtoseconds, whereas \(\tau_2\) represents the slow decay based on interband recombination and mid-gap trap capture \cite{keller_semiconductor_1996}.

The nonlinear reflectivity data is fitted using a model function of a two-level rate equation given by \cite{haiml_optical_2004}:

\begin{equation}
R^{\text{FlatTop}}(F) = R_{\text{ns}} \frac{\ln\left(1+\frac{R_{\text{lin}}}{R_{\text{ns}}} \left(\exp\left(\frac{F}{F_{\text{Sat}}}\right) - 1\right)\right)}{\frac{F}{F_{\text{Sat}}}} \cdot \exp\left(-\frac{F}{F_2}\right)%
\label{eq:one}
\end{equation}
with \( F_{\text{Sat}} \) as the saturation fluence, \( R_{\text{lin}} \) as the unsaturated reflectivity in the linear regime, \( R_{\text{ns}} \) as the highest reflectivity limited by nonsaturable losses, and \( F_2 \) as the pulse-duration dependent rollover parameter. The equation has an additional modification for a Gaussian beam profile:
\begin{eqnarray}
R^{\text{Gauss}}(F_p) = \frac{1}{2F_p} \int_{0}^{2F_p} R^{\text{FlatTop}}(F) \, dF
\label{eq:two}.
\end{eqnarray}
The pulse fluence \( F_P = \frac{E_P}{\pi w^2} \) is calculated from the pulse energy and the beam waist \( w \). 

\phantomsection
\vspace{0.7cm}

\begin{table*}
\caption{\label{tab:table1} Overview of SESAM parameters including the nonlinear characteristics.}
\begin{ruledtabular}
\begin{tabular}{cccccccccccc}
 \# SESAM & QW material & QW thickness & Barrier Material & \si{\lambda}$_{probe}$ & F$_{sat}$ & $\Delta R$ & $\Delta R_{ns}$ & F$_{2}$ & A & $\tau_1$ & $\tau_2$ \\
   &  & / \si{\nano\meter} & & / \si{\micro\meter} & / \si{\micro\joule\per\square\centi\meter} & / \% & / \% & / \si{\milli\joule\per\square\centi\meter} &  & / ps  & / ps \\ \hline
  \\
  1 & GaSb$_{0.92}$Bi$_{0.08}$ & 10 & GaSb & 2.06 & 1.1 & 1.4 & 0.5 & 3.5 & 0.25 & 0.29 & 6.0 \\
  2 & GaSb$_{0.92}$Bi$_{0.08}$ & 10 & AlAs$_{0.08}$Sb$_{0.92}$ & 2.06 & 1.2 & 1.8 & 4.4 & 5.1 & 0.54 & 0.65 & 29.8 \\
  3 & GaSb$_{0.87}$Bi$_{0.13}$ & 10 & GaSb & 2.3 & 1.9 & 0.6 & 1.2 & 4.2 & 0.96 & 0.12 & 3.0 \\
  SESAM from \cite{gaulke_optically_2024} & GaIn$_{0.27}$Sb$_{0.73}$ & 11.5 & GaSb & 2.06 & 0.6 & 2.3 & 0.7 & 2.5 & 0.76 & 0.21 & 8.9  \\
\end{tabular}
\end{ruledtabular}
\end{table*}

Table \ref{tab:table1} presents the full characterization data of the three GaSbBi-based SESAMs designed for operation in the \si{2} – \SI{2.3}{\micro\meter} wavelength range. 

\begin{figure}
\includegraphics[width=1.0\linewidth]{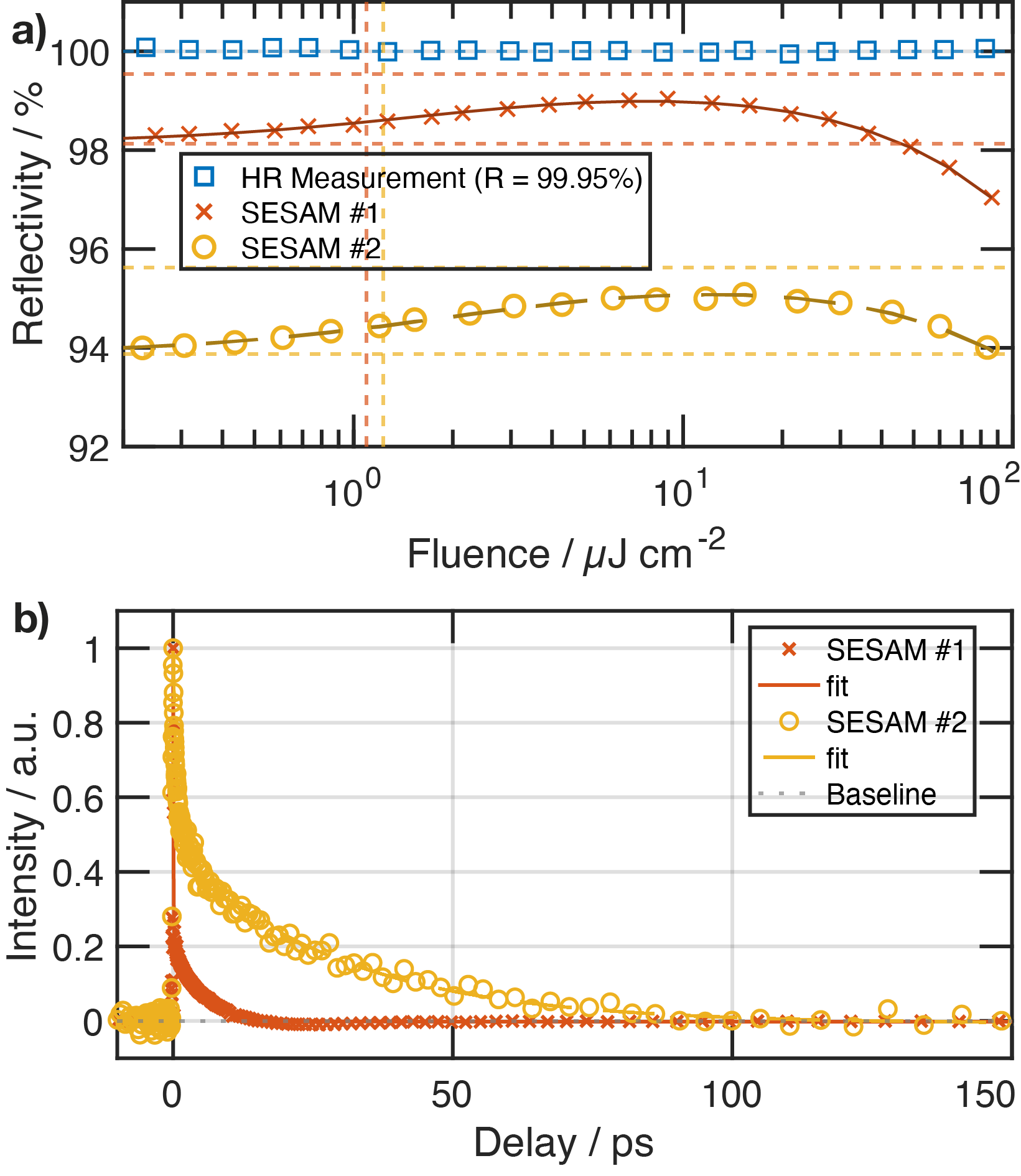}
\caption{\label{fig:Figure2} a) Nonlinear reflectivity measurement (dots) and fit (lines) for the two different SESAM \#1 (orange) and \#2 (yellow). b) Recovery time of SESAM \#1 and \#2, which shows that replacing the GaSb barrier with an AlAs\textsubscript{0.08}Sb\textsubscript{0.92} barrier results in a slightly longer recovery time.
}
\end{figure}

Figure \ref{fig:Figure2} presents the measured data including the fits of SESAM \#1 and \#2. SESAM \#1 exhibits a saturation fluence of \SI{1.09}{\micro\joule\per\square\centi\meter}, a modulation depth of \SI{1.41}{\percent}, and a rollover parameter \( F_2 \) of \SI{3.45}{\milli\joule\per\square\centi\meter}. Despite the effect of AR coating enhancing the modulation depth, this parameter defined per QW remains low. However, this does not pose a general problem as it can be compensated by increasing the number of QWs. For example, a three QW InGaSb/GaSb SESAM with a modulation depth of only \SI{0.8}{\percent} has been used for successful soliton modelocking at \SI{2.4}{\micro\meter} \cite{barh_watt-level_2021}. Moreover, SESAM \#1 exhibited nonsaturable losses of \SI{0.46}{\percent}. The recovery time measurements for SESAM \#1 revealed an intraband thermalization time \( \tau_1 \) of \SI{0.29}{\pico\second} and an interband relaxation time \( \tau_2 \) of \SI{6.03}{\pico\second}. Recovery times below tens of picoseconds are ideal for SESAM modelocking \cite{sieber_experimentally_2013}, while recovery times below \SI{50}{\pico\second} can still be used for soliton modelocking \cite{jung_experimental_1995}. Table \ref{tab:table1} also presents the characteristics of a standard AR-coated \SI{2}{\micro\meter} GaInSb/GaSb SESAM used for successful modelocking in \cite{gaulke_optically_2024}. Though this SESAM exhibits higher modulation depth and therefore slightly lower saturation fluence, the overall nonlinear reflectivity and temporal response characteristics closely mirror those of SESAM \#1, specifically the nonsaturable losses and the slow recovery time \( \tau_2 \). 

For SESAM \#2, the goal was to examine how the GaSbBi QWs behave in terms of the nonlinear SESAM characteristics when embedded in AlAs\textsubscript{0.08}Sb\textsubscript{0.92} barriers. This design exhibits a saturation fluence of \SI{1.23}{\micro\joule\per\square\centi\meter}, a modulation depth of \SI{1.75}{\percent}, and an increased \( F_2 \) parameter of \SI{5.09}{\milli\joule\per\square\centi\meter}. The increased modulation depth compared to SESAM \#1 can be attributed to the enhanced QW confinement provided by the AlAs\textsubscript{0.08}Sb\textsubscript{0.92} barrier. As expected, there is an increase in the \( F_2 \) parameter, validating our assumption that the use of AlAs\textsubscript{0.08}Sb\textsubscript{0.92} results in a lower TPA. To compare this with GaInSb QW SESAMs while also accounting for the AR coating, we assume that the rollover parameter depends solely on the second-order process of TPA. Using this assumption, we can calculate the ratio \( \frac{F_{2,\text{SESAM}\#2}}{F_{2,\text{SESAM}\#1}} \) to be approximately 1.5. A similar ratio is obtained when moving from GaInSb/GaSb to GaInSb/AlAs\textsubscript{0.08}Sb\textsubscript{0.92} SESAMs at \SI{2}{\micro\meter} \cite{schuchter_composition-controlled_2025}. This increase aligns closely with the GaSbBi-based SESAMs presented here, confirming our expectations.

However, the nonsaturable losses of SESAM \#2 are very high at \SI{4.37}{\percent}. We speculate that the high nonsaturable losses are primarily due to the unknown quality of the interfaces of the AlAs\textsubscript{0.08}Sb\textsubscript{0.92}/GaSbBi heterostructure as well as possible detrimental effects related to the residual background As-pressure. This resulted from the growth of the AlAs\textsubscript{0.08}Sb\textsubscript{0.92} layers. This has a strong influence on Bi incorporation, possibly promoting Bi droplet formation \cite{hilska_epitaxial_2019}.

\begin{figure}
\includegraphics[width=1\linewidth]{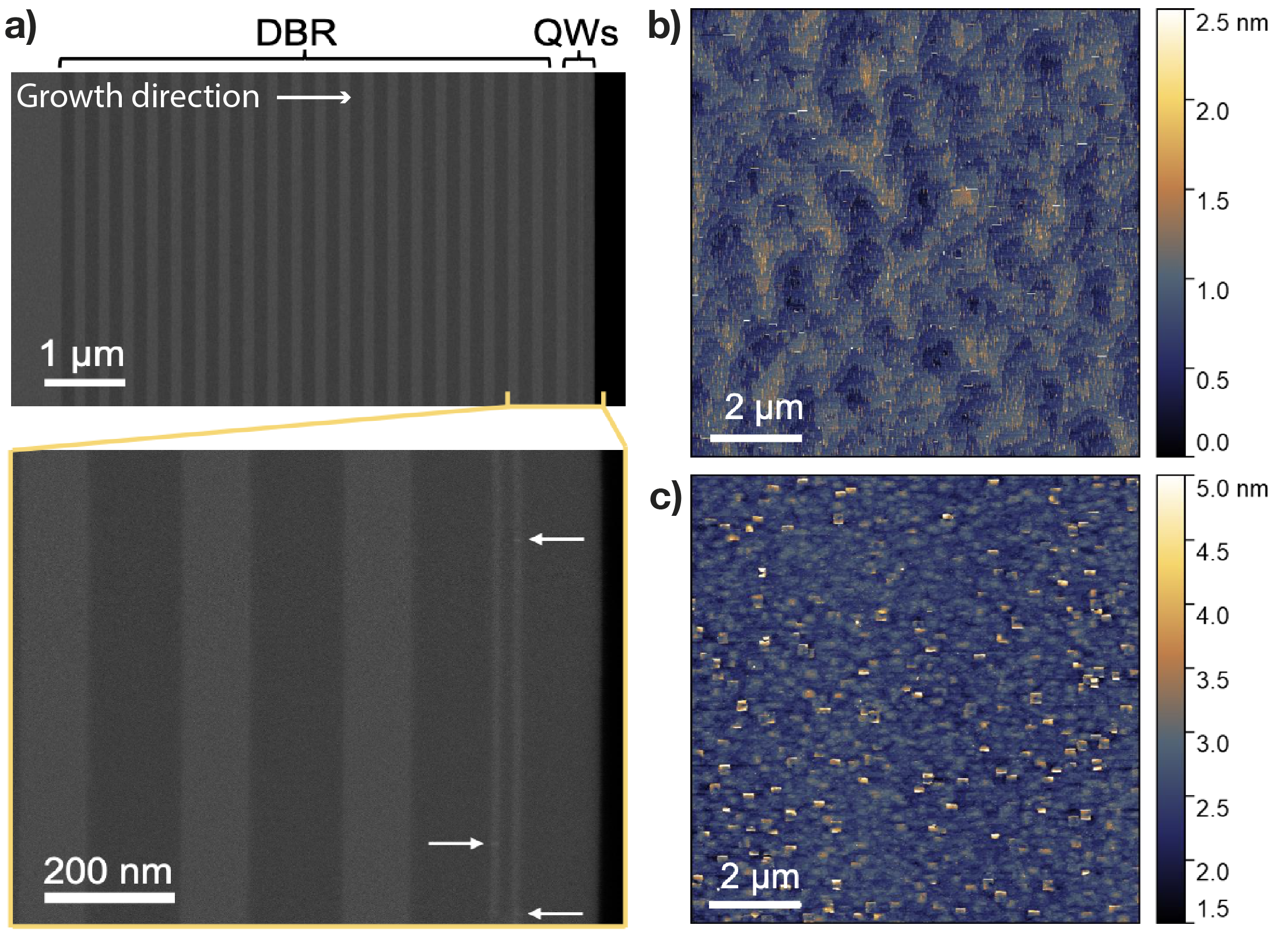}
\caption{\label{fig:Figure3} a) Cross-sectional back-scattered-electron detector SEM image of SESAM \#2. The top image shows the full structure. The bottom zooms in on the QW region, revealing bright spots in the QWs noted by arrows. b-c) 10$\times$10 \si{\micro\meter\squared} AFM scans of the as-grown surface of (b) SESAM \#1 and (c) SESAM \#2.}
\end{figure}

To investigate this, we characterized the as-grown structure with both cross-sectional scanning electron microscopy (SEM) and atomic force microscopy (AFM). Figure \ref{fig:Figure3}a) shows backscattered electron contrast SEM images of SESAM \#2, demonstrating excellent DBR growth but also revealing defects in the QWs as seen by bright spots. This brighter contrast compared to the surrounding material implies that the defects have a higher fraction of atoms with a higher atomic number, such as Bi. In fact, Bi is well known to either (i) nucleate on the growth front as metallic droplets \cite{hilska_epitaxial_2019, yue_molecular_2018}, (ii) form varying composition modulations \cite{wu_detecting_2015}, or (iii) cluster via post-growth annealing \cite{straubinger_thermally_2018}. For example, Bi droplets could have formed due to the influence of residual As-pressure in the MBE from the AlAs\textsubscript{0.08}Sb\textsubscript{0.92} barrier growth between the two QWs and could have also resulted in a slightly varying III/V ratio for each of the QWs. If droplets formed, the subsequent overgrowth of material could have induced rougher interfaces, leading to higher nonsaturable losses.

AFM images taken from the SESAM \#1 and SESAM \#2 surfaces can be seen in Figures \ref{fig:Figure3}b) and \ref{fig:Figure3}c), respectively. SESAM \#1 exhibits a smooth epitaxial atomic terrace structure of the top-most GaSb layers in the as-grown material. The root-mean-square (RMS) roughness is measured to be \SI{0.270}{\nano\meter} for the 10$\times$10 \si{\micro\meter\squared} AFM image. Closer inspection of the surface of SESAM \#1 reveals nanometer-sized deposits on the surface, which lie along the major crystalline axes. These small deposits are due to unintended Sb condensation on the surface during the post-growth cooling procedure \cite{arpapay_redundant_2014}. Due to their small size, they likely do not influence the nonsaturable losses. This is corroborated by the same cooling procedure used in the growth of standard GaInSb/GaSb SESAMs, which exhibited low \( R_{\text{ns}} \).

In contrast to SESAM \#1, SESAM \#2 with the AlAs\textsubscript{0.08}Sb\textsubscript{0.92} barriers exhibits a generally rougher surface morphology, as shown in Figure \ref{fig:Figure3}c). The RMS surface roughness is correspondingly increased to a value of \SI{0.378}{\nano\meter} for the 10$\times$10 \si{\micro\meter\squared} scan area. While the RMS roughness is still relatively low, the morphology does not show a clear terrace structure but rather a distinctly mounded surface structure with square defects. These square defects and the mounded surface morphology could be the result of unintended droplet formation during the growth of the AlAs\textsubscript{0.08}Sb\textsubscript{0.92}/GaSbBi QW heterostructure and subsequent overgrowth of the QWs with AlAs\textsubscript{0.08}Sb\textsubscript{0.92} at higher temperature. Overall, the SEM and AFM structural analysis from SESAM \#2 indicates that the growth parameters or the growth sequence of the AlAs\textsubscript{0.08}Sb\textsubscript{0.92}/GaSbBi cavity structure was suboptimal, resulting in much larger scattering-related nonsaturable losses. 

The absorber recovery dynamics of SESAM \# 2 were also measured and can be seen in Figure \ref{fig:Figure2}b) (yellow data points). We want to reiterate that previous studies \cite{schuchter_composition-controlled_2025} have reported extremely slow recovery times (\textgreater \SI{500}{\pico\second}) for GaInSb/AlAs\textsubscript{0.08}Sb\textsubscript{0.92} SESAMs, making them unsuitable for modelocking applications. The GaSbBi/AlAs\textsubscript{0.08}Sb\textsubscript{0.92} SESAM \#2, on the other hand, exhibits a fast interband relaxation time \( \tau_2 \) of \SI{29.8}{\pico\second}, which clearly deviates from the trend of the standard GaInSb/AlAs\textsubscript{0.08}Sb\textsubscript{0.92} SESAM behavior. Although slightly slower than SESAM \#1, this recovery time remains suitable for modelocking \cite{jung_experimental_1995}. This finding suggests that the underlying decay mechanisms in GaSbBi QWs differ from those in GaInSb QWs. The AlAs\textsubscript{0.08}Sb\textsubscript{0.92} barrier does not dominate the recovery behavior but only introduces a slight increase in recovery time, which should not prevent modelocking \cite{jung_experimental_1995}. The GaSbBi/AlAs\textsubscript{0.08}Sb\textsubscript{0.92} SESAM \#2 exhibits both a high \( F_2 \) parameter and fast recovery dynamics, and with further growth optimization to decrease nonsaturable losses, could serve as an ideal SESAM for high-fluence modelocked operation. 

To explore GaSbBi QWs at longer wavelengths, SESAM \#3 was designed for operation at \SI{2.3}{\micro\meter} by increasing the Bi content to \SI{13}{\percent}, showcasing the flexibility of GaSbBi growth for SWIR applications. SESAM \#3 exhibits a saturation fluence of \SI{1.77}{\micro\joule\per\square\centi\meter}, a modulation depth of \SI{0.57}{\percent}, nonsaturable losses of \SI{1.2}{\percent}, and a rollover parameter of \SI{4.22}{\milli\joule\per\square\centi\meter}. It also demonstrates excellent recovery dynamics, with an intraband thermalization time of \SI{0.12}{\pico\second} and an interband relaxation time of \SI{2.95}{\pico\second}. It is worth noting that the nonsaturable losses have slightly increased from SESAM \#1 to SESAM \#3, reaching \SI{1.2}{\percent}. This increase is attributed to the difficulty in growing high Bi-content GaSb\textsubscript{0.87}Bi\textsubscript{0.13}, which has a more narrow growth parameter window \cite{hilska_epitaxial_2019}. To achieve the high Bi content (\SI{13}{\percent}), the growth temperature had to be reduced from \SI{315}{\celsius} to \SI{295}{\celsius}, further influencing the material quality. Nonetheless, the recovery dynamics are well suited for modelocking and highlight the potential of GaSbBi for higher-wavelength SESAM applications.

\phantomsection
\vspace{0.7cm}

In summary, the fabrication and characterization of SWIR SESAMs utilizing GaSbBi QWs for operation in the \si{2} – \SI{2.4}{\micro\meter} wavelength range was reported for the first time. Three SESAMs, each incorporating GaSbBi QWs, were grown in different configurations to mitigate performance challenges inherent to standard GaSb-based SESAMs. 

SESAM \#1 utilized GaSbBi/GaSb QWs designed for \SI{2.06}{\micro\meter}, which possessed characteristics similar to conventional GaInSb/GaSb SESAMs, including fast recovery dynamics (\textless \SI{10}{\pico\second}) and low nonsaturable losses (\SI{0.46}{\percent}), with an early reflectivity rollover related to two-photon absorption (TPA) in GaSb barriers. 

SESAM \#2 used AlAs\textsubscript{0.08}Sb\textsubscript{0.92} as the barrier material, which successfully reduced TPA resulting in an increased rollover parameter (\(F_2\)), but with significantly increased nonsaturable losses ascribed to a non-optimized growth procedure for the GaSbBi/AlAs\textsubscript{0.08}Sb\textsubscript{0.92} heterostructure. Furthermore, SESAM \#2 maintained its favorable recovery time (\textless \SI{30}{\pico\second}), highlighting the potential of GaSbBi QWs to overcome the longer recovery time challenge associated with AlAs\textsubscript{0.08}Sb\textsubscript{0.92} barriers in traditional GaInSb-based designs.

For SESAM \#3, the operational wavelength was extended to \SI{2.3}{\micro\meter} by increasing the Bi content in the QWs while otherwise following the design of SESAM \#1. While nonsaturable losses increased slightly due to the challenges of incorporating higher Bi content while maintaining material quality, the recovery dynamics remained fast (\textless \SI{10}{\pico\second}) and well-suited for modelocking. 

These results underscore the promise of GaSbBi QWs as a versatile material platform for SESAMs, offering solutions to key limitations such as TPA, strain limitations, and slow recovery times induced by AlAs\textsubscript{0.08}Sb\textsubscript{0.92} barriers in traditional SESAM designs. 
For increased modulation depth, we can simply increase the number of QWs to three, as similarly done in \cite{barh_watt-level_2021}. However, further optimization of growth techniques of GaSbBi QWs to reduce nonsaturable losses is still required, especially for barrier materials containing Al. Nonetheless, GaSbBi-based SESAMs could become a versatile tool for ultrafast laser technologies across the mid-IR spectrum.

\begin{acknowledgments}
The authors would like to acknowledge the suport received from FIRST clean room facility at ETH Zurich. They acknowledge helpful discussion with Nicolas Huwyler and Matthias Golling. The authors also acknowledge Metin Patli for providing the AFM scans. The SESAM fabrication work was financially supported by the Academy of Finland Flagship program PREIN (decision 320168) and the H2020 European Research Council (ERC advanced grant 787097 ONE-MIX).
\end{acknowledgments}

\section*{Data Availability Statement}
The data that support the findings of this study are available from the corresponding author upon reasonable request. 

\section*{References}
\bibliography{aipsamp}

\end{document}